\documentclass[12pt]{article}
\usepackage{graphicx}
\usepackage{amsmath, amsthm, amssymb}

\newcommand\pubnumber{LHCb-PROC-2011-001}
\newcommand\pubdate{\today}

\def\CERN{Organisation Europ\'{e}enne pour la Recherche Nucl\'{e}aire\\ 
Geneva, CH-1211, Switzerland}

\def\Title#1{\begin{center} {\Large #1 } \end{center}}
\def\Author#1{\begin{center}{ \sc #1} \end{center}}
\def\Address#1{\begin{center}{ \it #1} \end{center}}

\def\DsK{$B^0_{s}\rightarrow D^\pm_{s}K^\mp$}
\def\DPi{$B^0_{d}\rightarrow D^\pm\pi^\mp$}
\def\DsPi{$B^0_{s}\rightarrow D^-_{s}\pi^+$}

\def\DsKone{$B^0_{s}\rightarrow D^\pm_{s}K^\mp_1$}
\def\DstMuNu{$B^{0}_{d}\rightarrow D^{*}\mu\nu$}
\newcommand\pubblock{\rightline{\begin{tabular}{l} \pubnumber\\
         \pubdate  \end{tabular}}}
\newenvironment{Abstract}{\begin{quotation}  }{\end{quotation}}
\newenvironment{Presented}{\begin{quotation} \begin{center} 
             PRESENTED AT\end{center}\bigskip 
      \begin{center}\begin{large}}{\end{large}\end{center} \end{quotation}}
\def\Acknowledgements{\bigskip  \bigskip \begin{center} \begin{large}
             \bf ACKNOWLEDGEMENTS \end{large}\end{center}}

\def\beq{\begin{equation}}
\def\eeq#1{\label{#1}\end{equation}}
\def\eeqn{\end{equation}}

\def\beqa{\begin{eqnarray}}
\def\eeqa#1{\label{#1}\end{eqnarray}}
\def\eeqan{\end{eqnarray}}

\let\bar=\overbar

\def\Dslash{\not{\hbox{\kern-4pt $D$}}}
\def\dslash{\not{\hbox{\kern-2pt $\del$}}}

\def\msb{{\bar{\ssstyle M \kern -1pt S}}}

\begin{document}
\begin{titlepage}
\pubblock

\vfill
\Title{Time-dependent measurements of the CKM angle $\gamma$ at LHCb}
\vfill
\Author{Vladimir V. Gligorov, on behalf of the LHCb collaboration}
\Address{\CERN}
\vfill
\begin{Abstract}
The startup of the LHC opens many new frontiers in precision flavour physics, in particular
expanding the field of precision time-dependent CP violation measurements to the $B^0_s$ system. This contribution reviews the
status of time-dependent measurements of the CKM angle $\gamma$ at the LHC's dedicated flavour physics
experiment, LHCb. Particular attention is given to the measurement of $\gamma$ from the decay mode \DsK,
a theoretically clean and precise method which is unique to LHCb. The performance of the LHCb detector 
for this and related modes is reviewed in light of early data taking and found to be close to the nominal
simulation performance, and the outlook for these measurements in 2011 is briefly touched on.
\end{Abstract}
\vfill
\begin{Presented}
CKM 2010\\
University of Warwick, United Kingdom
\end{Presented}
\vfill
\end{titlepage}
\def\thefootnote{\fnsymbol{footnote}}
\setcounter{footnote}{0}
\section{Introduction}
At the dawn of the LHC era, the CKM angle $\gamma$ remains the least well measured component of the Unitarity Triangle which describes CP violation
in the Standard Model.
The current status of $\gamma$ measurements and the motivation for measuring $\gamma$ precisely are both discussed 
elsewhere in these proceedings~\cite{Zupan,Anton}.
This document will outline the current status of time-dependent $\gamma$
measurements at the LHCb experiment. One of the unique things about LHCb is
its simultaneous access to large quantities of $B^0_s$ decays and ability to resolve $B^0_s$ oscillations. In particular, LHCb can measure
$\gamma$ from the interference in mixing and decay of the mode \DsK. This is a theoretically pristine~\cite{Aleksan:1991nh,Fleischer:2003yb}
measurement of $\gamma$, but has hitherto been experimentally inaccesible.
Due to the limited space, the focus here will be on this measurement,
but other time-dependent measurements are also being pursued at LHCb~\cite{Akiba:2008zza}. 
\section{Measuring $\gamma$ from \DsK}
Before discussing the current experimental prospects, it is
worth reminding ourselves of the phenomenology of this decay. The treatment here (and the LHCb measurement strategy)
follows the notation and method introduced in~\cite{Fleischer:2003yb}.

The time-dependent CP asymmetry in the decay $B^0_q\rightarrow D_q \overline{u}_q$ is given by
\begin{equation}
\label{eq:rateasymm}
A\left(t\right) =  
\frac{\frac{1-|\xi_q|^2}{1+|\xi_q|^2}\cos\left(\Delta M_qt\right)+
\frac{2\textrm{Im}\left(\xi_q\right)}{1+|\xi_q|^2}\sin\left(\Delta M_qt\right)}
{\cosh\left(\Delta\Gamma_qt/2\right)-\frac{2\textrm{Re}\left(\xi_q\right)}{1+|\xi_q|^2}\sinh\left(\Delta\Gamma_qt/2\right)},
\end{equation}
where $q$ stands for the $d$ or $s$ quark, and $u_q$ is a pion or kaon. $\Delta M_q$ is the mass difference
between the $B_{L}$ and $B_{H}$ eigenstates in the $B_q$ system, and $\Delta\Gamma_q$ the decay
width difference between the ``heavy'' and ``light'' $B_q$ mass eigenstates. The sensitivity to $\gamma$ comes through
\begin{equation}
\label{eq:xi}
\xi_q = -\left(-1\right)^Le^{-i\left(\phi_q+\gamma\right)}\left[\frac{1}{x_q e^{i\delta_q}}\right],
\end{equation}
where $\phi_q$ is the mixing phase between $B^0_q$ and $\overline{B}^0_q$, $\delta_q$ is a strong
phase difference between the decay channels $\overline{B^0_q}\rightarrow D_q \overline{u}_q$ and
$B^0_q\rightarrow D_q \overline{u}_q$, which is not CP-violating, and $x_q$ represents the level of interference
between the $B^0_q$ and $\overline{B}^0_q$ decaying into the same final state.

Current time-dependent measurements~\cite{Rubin,Onuki} of $\gamma$ 
from $B^0_d$ decays suffer from the small values of $\Delta\Gamma_d\approx 0$ and $x_d\approx 0.04$, which mean that the bottom and top left terms
in Equation~\ref{eq:rateasymm} approximate to 1, leaving only the ``sinusoidal'' observables in the top right term to constrain $\gamma$. It can be shown
that this results in an eightfold ambiguity, and as well means that the value of $x_d$, computed from SU(3) arguments, must be an external input to the fit.
On the other hand $\Delta\Gamma_s$ could be sizeable, while $x_s\approx 0.4$, so that in the $B^0_s$ system there are three
extra observables. This both reduces the ambiguity to a twofold one, because of the ``hyperbolic'' observables in the bottom term,
and allows $x_s$ to be fitted for directly, eliminating any SU(3) uncertainties in the measured value of $\gamma$. 
The measurement of $\gamma$ from \DsK benefits from a simultaneous fit with \DsPi, which allows both $\Delta m_s$ and the flavour tagging power to be fitted for directly.
Taken together these facts make the extraction of $\gamma$ from \DsK competitive with time-integrated methods~\cite{Williams}; previous simulation
studies~\cite{Akiba:2008zza} have indicated a precision of $10^\circ$ with 2~fb$^{-1}$ of integrated luminosity.
\section{Prospects for $\gamma$ from \DsK at LHCb}
During 2010 LHCb has collected $\approx 37$~pb$^{-1}$ of data, sufficient for a first comparison with the expected~\cite{Akiba:2008zza} Monte Carlo simulation (MC) performance
in these decay modes.
At the time of the CKM conference only $\approx 3$~pb$^{-1}$ had been collected; the results shown here therefore supersede the related conference talk.
\subsection{Selection performance}
\begin{figure}[htb]
\centering
\includegraphics[height=1.5in]{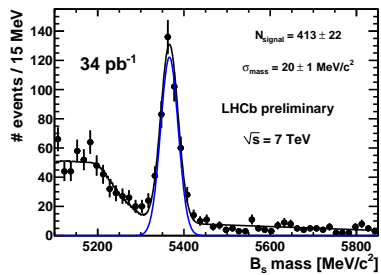}
\caption{Selected \DsPi events, in which the $D_s$ decays to $\phi\pi$. The total yield including other resonant and non-resonant $D_s$ decay modes
is $\approx2$ times larger.}
\label{fig:mphipi}
\end{figure}

The offline selection performance in this family of modes is illustrated in Figure~\ref{fig:mphipi} for \DsPi. The offline purity and reconstruction efficiency
are broadly in line with MC expectations, however the overall yield per~pb$^{-1}$ is approximately a factor two lower than expected. This is because
of the running conditions encountered in October of 2010: a high number of $pp$ interactions in each bunch crossing and a low number of filled bunches.
LHCb was designed for precisely the opposite running conditions and as a result strong event multiplicity cuts had to be applied at the trigger level, leading to
a decrease in efficiency.
\subsection{Propertime resolution and flavour tagging}
\begin{figure}[htb]
\centering
\includegraphics[height=1.5in]{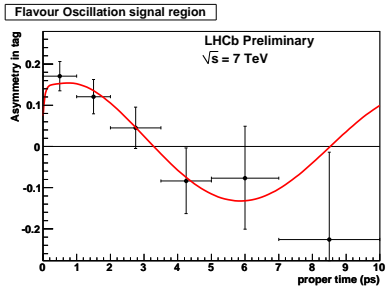}
\caption{Oscillations in \DstMuNu with $\approx 2$~pb$^{-1}$ of data. The value of $\Delta m_d$ from an unconstrained maximum likelihood fit is $(0.53 \pm 0.08) \times 10^{12}$~$\hbar s^{-1}$. }
\label{fig:B0Asym}
\end{figure}

The LHCb propertime resolution is currently around 50~fs, close to the MC expectation of 40~fs. Flavour tagging at LHCb includes opposite side tagging, which infers the 
production flavour of the $B^0_{d,s}$ meson by tagging the ``other'' B decay in the event, and same side tagging, which primarily tags the production flavour of a $B^0_s$
meson by searching for the associated charged kaon produced in the fragmentation. The opposite side flavour tagging performance is around $60\%$ of the
nominal MC expectation, while the same side tagging performance is still under evaluation.
These issues are discussed in more detail
elsewhere in these proceedings~\cite{Menzemer}. A degradation in tagging power or propertime resolution results in reduced sensitivity to $\gamma$ per~pb$^{-1}$ of data,
but both are well within the limits required to make time-dependent CP measurements at LHCb intrinsically viable. Indeed LHCb has already observed $B^0_d$ oscillations, as
seen in Figure~\ref{fig:B0Asym}.
\section{\DsK and ambiguities: a closer look}
\begin{figure}[ht]
\centering
\includegraphics[height=2.5in]{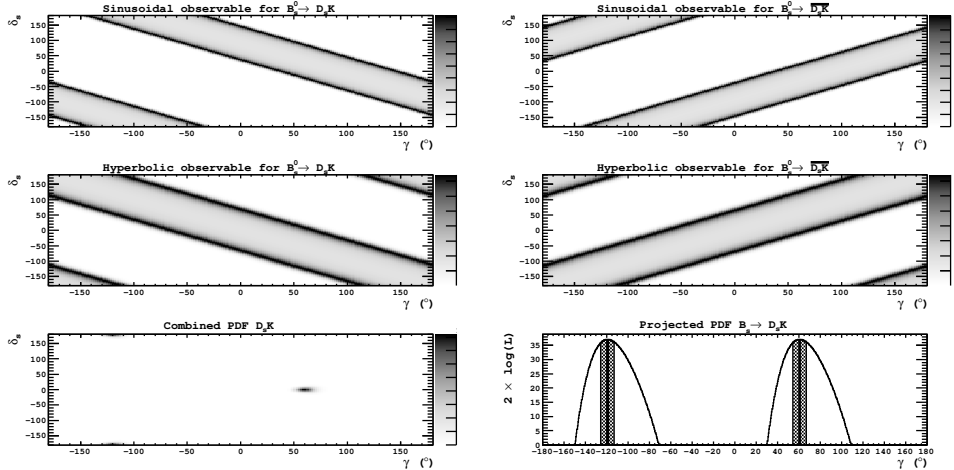}
\caption{The top four plots show the likelihoods of the \DsK~CP observables with 31k signal events and LHCb MC performance~\cite{Akiba:2008zza}.
The input values of $\gamma$, $\delta_s$, and $\frac{\Delta\Gamma_s}{\Gamma_s}$ are $60^\circ$, $10^\circ$, and $10\%$. The bottom left plot
shows the combined likelihood in the $\gamma-\delta_s$ plane and the bottom right the projection onto $\gamma$, where the thatched area is the 1$\sigma$
region and the dark vertical line the central value. All but two of
the ambiguities are excluded and the central value of $\gamma$ is unbiased.}
\label{fig:gammaFromDsK_10fb}
\end{figure}
\begin{figure}[ht]
\centering
\includegraphics[height=2.5in]{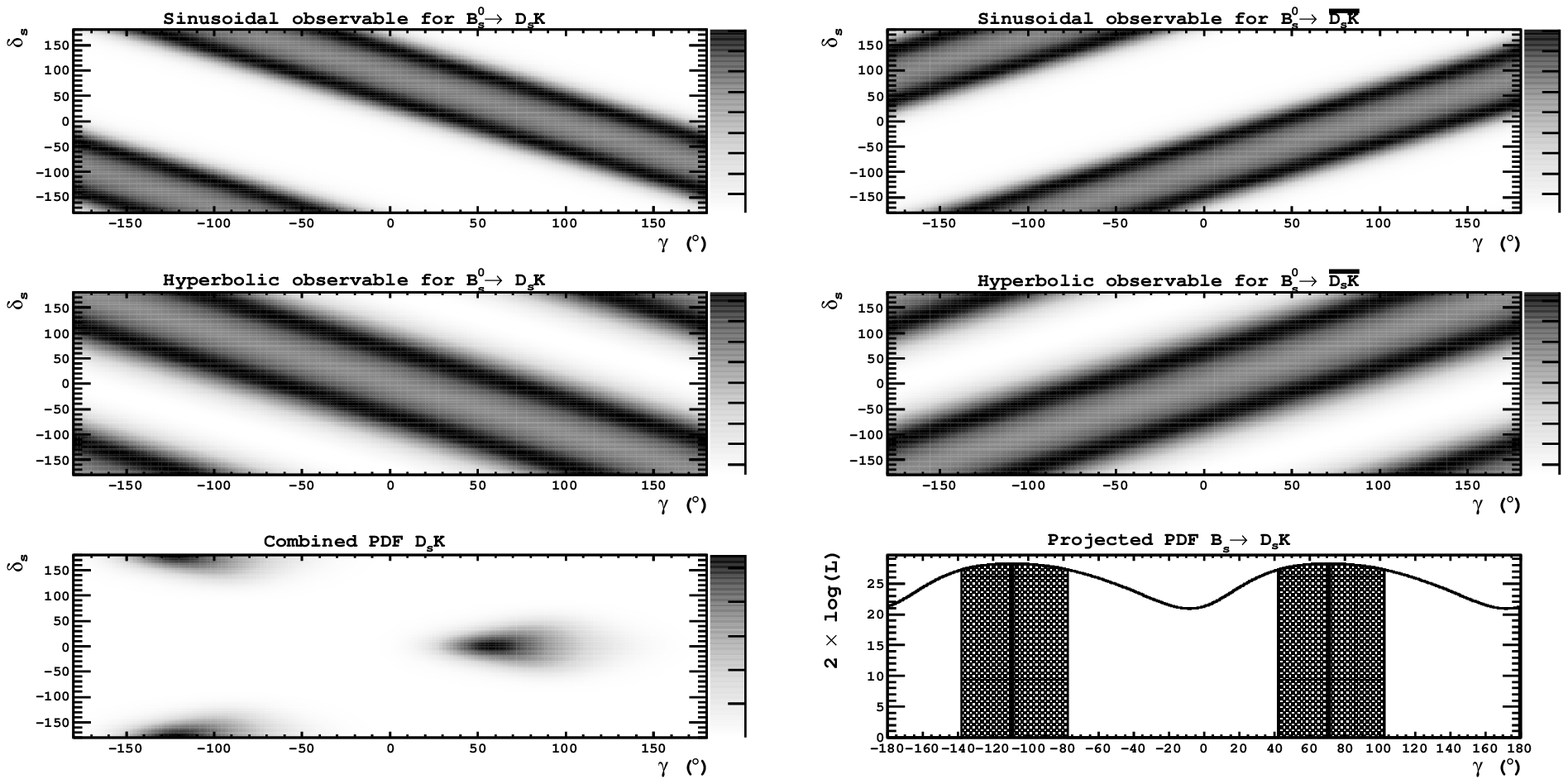}
\caption{As Figure~\ref{fig:gammaFromDsK_10fb} but for 1.55k signal events. The ambiguities
blend into two blocks of four degenerate solutions each, and the central value of $\gamma$ is biased.}
\label{fig:gammaFromDsK_500pb}
\end{figure}
As was already remarked, the measurement of $\gamma$ from \DsK suffers from only a twofold ambiguity in the case that $\frac{\Delta\Gamma_s}{\Gamma_s}$ is
sizeable (say around $10\%$), but this only holds in the limit of sufficient statistics. Figures~\ref{fig:gammaFromDsK_10fb}~and~\ref{fig:gammaFromDsK_500pb} illustrate this point.
Note that, unlike in the $B^0_{d}$ system where the sensitivity to $\gamma$ shows a significant dependence on the size of the strong phase $\delta_d$,
this result is independent of the size of the strong phase $\delta_s$. The figures illustrate how the extra hyperbolic observables present in the $B^0_s$ system
remove ambiguities: their likelihood in the $\delta_s-\gamma$ plane is staggered compared to the sinusoidal observables.
It also shows that the \DsK measurement requires a certain ``critical mass'' of signal events below which there is a significant bias in the measured
value of $\gamma$. In this case a combined fit with the decay modes $B^{0}_d\rightarrow D^{\pm\left(*\right)}\pi^\mp$ under the assumption of U-spin symmetry~\cite{Fleischer:2003yb,Gligorov:2008zzb} might allow for the ambiguous solutions to be partially resolved.
\section{Conclusions and outlook}
The long term outlook for time-dependent $\gamma$ measurements at LHCb is excellent: LHCb has already very nearly
achieved its nominal MC selection efficiency and propertime resolution, and $B^0_d$ oscillations have been observed. 
The ultimate systematic limitations will not be clear until the measurement of $\gamma$ from \DsK has been performed, but there is no reason to suspect
that any crippling problems lurk around the corner. 
On the other hand, the short term outlook is more uncertain. The LHC plans a long shutdown in 2012 in order to increase the beam energy from the present $3.5$~TeV to the
nominal 7~TeV, and it is natural to ask what precision LHCb can achieve on $\gamma$ with the 2011 dataset, expected to be around 1~fb$^{-1}$.
Extrapolating from the current yield of
approximately 25 \DsPi events per pb$^{-1}$, and adjusting for the relative yields of these modes as measured in MC studies, LHCb could expect around 2100 \DsK decays in 2011.
This will be enough to measure the CP observables, but not enough for
an unambiguous measurement of $\gamma$. Of course, improvements are being worked on: higher trigger efficiencies,
the inclusion of suppressed $D_s$ decays, or the inclusion of related decay modes such as \DsKone in the fit. Much also depends
on the poorly known branching ratio of \DsK, but around 4k events will be needed to have a precise measurement in 2011.
\Acknowledgements
I am grateful to the organizers of CKM 2010 for their invitation, and to
Robert Fleischer and Stefania Ricciardi for many productive discussions
about $\gamma$.


\begin{thebibliography}{99}
\bibitem{Zupan}
J. Zupan, these proceedings.
\bibitem{Anton}
A. Poluektov, these proceedings.
\bibitem{Aleksan:1991nh}
R.~Aleksan, I.~Dunietz and B.~Kayser,
Z.\ Phys.\  C {\bf 54} (1992) 653.
\bibitem{Fleischer:2003yb}
R.~Fleischer,
Nucl.\ Phys.\  B {\bf 671} (2003) 459
[arXiv:hep-ph/0304027].
\bibitem{Akiba:2008zza}
K.~Akiba {\it et al.},
``Determination of the CKM-angle gamma with tree-level processes at LHCb,''
\bibitem{Rubin}
A. Rubin, these proceedings.
\bibitem{Onuki}
Y. Onuki, these proceedings.
\bibitem{Williams}
M. Williams, these proceedings.
\bibitem{Menzemer}
S. Hansmann-Menzemer, these proceedings.
\bibitem{Gligorov:2008zzb}
V.~Gligorov and G.~Wilkinson,
``Strategies for measuring gamma from the decay channels \DPi and \DsK at LHCb,''
\end{thebibliography}
\end{document}